\documentclass{PoS}

\newcommand{\nijmegen}{Radboud Univ.\ Nijmegen}
\newcommand{\astron}{ASTRON}
\newcommand{\karlsruhe}{KIT}
\newcommand{\erlangen}{Univ.\ of Erlangen-Nuremberg}
\newcommand{\groningen}{Univ.\ of Groningen}
\newcommand{\manchester}{Univ.\ of Manchester}
\newcommand{\adelaide}{Univ.\ of Adelaide}
\newcommand{\iowa}{Univ.\ of Iowa}
\newcommand{\atnf}{CSIRO ATNF}
\newcommand{\lebedev}{Lebedev Physical Institute}
\newcommand{\santiago}{Univ.\ de Santiago de Compostela}
\newcommand{\brussels}{Vrije Univ.\ Brussel}

\title{The lunar Askaryan technique with the Square Kilometre Array}

\ShortTitle{The lunar technique with SKA}

\usepackage[authoryear,sort]{natbib}
\bibpunct{(}{)}{;}{a}{}{,}

\author{
\speaker{C.~W.~James}$^1$, J.~Alvarez-Mu\~niz$^2$, {J.D.~Bray}$^3$, S.~Buitink$^4$, R.D.~Dagkesamanskii$^5$, R.D.~Ekers$^6$, H.~Falcke$^{7,8}$, K.G.~Gayley$^9$, T.~Huege$^{10}$, M.~Mevius$^{11}$, R.L.~Mutel$^9$, R.J.~Protheroe$^{12}$, O.~Scholten$^{4,11}$, R.E.~Spencer$^{3}$, and S.~ter~Veen$^7$ \\
 $^1$\erlangen ;
 $^2$\santiago ;
 $^{3}$\manchester ;
$^{4}$\brussels
 $^5$\lebedev ;
 $^6$\atnf ;
 $^7$\nijmegen ;
 $^8$\astron ;
 $^9$\iowa ;
 $^{10}$\karlsruhe ;
 $^{11}$\groningen ;
 $^{12}$\adelaide \\
E-mail: \email{clancy.james@physik.uni-erlangen.de}
}

\abstract{The lunar Askaryan technique is a method to study the highest-energy cosmic rays, and their predicted counterparts, the ultra-high-energy neutrinos. By observing the Moon with a radio telescope, and searching for the characteristic nanosecond-scale Askaryan pulses emitted when a high-energy particle interacts in the outer layers of the Moon, the visible lunar surface can be used as a detection area. Several previous experiments, at Parkes, Goldstone, Kalyazin, Westerbork, the ATCA, Lovell, LOFAR, and the VLA, have developed the necessary techniques to search for these pulses, but existing instruments have lacked the necessary sensitivity to detect the known flux of cosmic rays from such a distance. This will change with the advent of the SKA.

The Square Kilometre Array (SKA) will be the world's most powerful radio telescope. To be built in southern Africa, Australia and New Zealand during the next decade, it will have an unsurpassed sensitivity over the key 100 MHz to few-GHZ band. We introduce a planned experiment to use the SKA to observe the highest-energy cosmic rays and, potentially, neutrinos. The estimated event rate will be presented, along with the predicted energy and directional resolution. Prospects for directional studies with phase 1 of the SKA will be discussed, as will the major technical challenges to be overcome to make full use of this powerful instrument. Finally, we show how phase 2 of the SKA could provide a vast increase in the number of detected cosmic rays at the highest energies, and thus to provide new insight into their spectrum and origin.}

\FullConference{The 34th International Cosmic Ray Conference,\\
		30 July- 6 August, 2015\\
		The Hague, The Netherlands}

\begin{document}

\section{Introduction}
\label{sec:intro}

The lunar Askaryan technique was proposed by \citet{dagkesamanskii1989} as a method to study the highest-energy cosmic rays and neutrinos. By observing the Moon from a satellite or ground-based telescope and searching for Askaryan emission \citep{askaryan1962} from particle interactions, the entire visible lunar surface (of order $20$ million km$^2$) can be utilised as a particle detector. This method has been used successfully to place limits on exotic (`top-down') models of ultra-high-energy (UHE; $E \gtrsim 10^{20}$~eV) neutrino production \citep{gorham2004a,buitink2010}. However, due to the large Earth-Moon distance ($3.844 \cdot 10^{5}$~km on average), the signals on Earth are very weak, which has precluded current radio telescopes from being sensitive to either neutrino fluxes from conventional `bottom-up' cosmic ray acceleration models, or to the cosmic ray flux itself.

This situation will change with the advent of the Square Kilometre Array\footnote{www.skatelescope.org} (SKA). The largest radio telescope ever conceived, the SKA will consist of multiple components, with the low-frequency telescope SKA-low covering the range $50$--$350$~MHz. This instrument, to be built in Western Australia \citep{SKA_baseline_design} between 2018--2023 (phase 1) and 2023--2030 (phase 2), is particularly appealing as an UHE particle detector, due to the increased angular width of the Askaryan emission at low frequencies \citep{NuMoonSim}.
When complete, SKA-low is expected to detect an unprecedented number of UHE cosmic rays, with sufficient angular resolution to study their origin.

In order to enable lunar Askaryan observations with SKA-low, groups that have performed observations with existing radio telescopes --- LUNASKA \citep{james2010,2015APh....65...22B}, NuMoon \citep{buitink2010}, RAMHAND \citep{beresnyak2005}, LaLUNA \citep{spencer2010}, and RESUN \citep{jaeger2010} --- have jointly\footnote{Together with a project to make precision measurements of extensive air showers --- see \citet{ska_eas}.} formed the SKA High Energy Cosmic Particles (HECP) Focus Group\footnote{http://astronomers.skatelescope.org/home/focus-groups/high-energy-cosmic-particles/}. This group will apply the necessary methods and technology, developed as part of the aforementioned forerunner projects, to observations with SKA-low, and work with SKA engineers to implement this observation mode. In parallel, simulations have been performed in order to estimate the sensitivity of SKA-low during both its initial phase $1$ deployment, and final phase $2$ configuration.

The results of this effort are reported here. Sections \ref{sec:ska} and \ref{sec:simulations} respectively give an overview of SKA-low and the simulation chain used to simulate its sensitivity to lunar Askaryan pulses. In Section \ref{sec:cosmic_rays}, estimates of the ability of SKA-low phases $1$ and $2$ to detect and study UHE cosmic rays (CR) are given, including preliminary energy and direction resolutions. Section \ref{sec:neutrinos} discusses the prospects to study UHE neutrinos and constrain the remaining top-down models of UHE particle production. Finally, Section \ref{sec:discussion} reviews the science case of the HECP (lunar) group in light of these results and in comparison to other experiments, and discusses the necessary technical requirements in order to achieve the simulated sensitivity.

\section{SKA-low}
\label{sec:ska}

The first phase of the low-frequency component of the SKA, SKA-low phase $1$, will consist of $130,000$ log-periodic dipole antennas deployed in $512$ stations of $256$ antennas each, giving a total collecting area of $0.4$~km$^2$ \citep{SKA_baseline_design}. Half the stations will be deployed in a dense core of less than $1$~km diameter, while the remainder will be located at distances of up to $65$~km. Dual polarisation signals covering the range $50$--$350$~MHz from all antennas in each station will be digitised and added in-phase to form `station beams' with full-width half maximum (FWHM) of $> 1.4^{\circ}$. The second phase of SKA-low will be deployed in a similar configuration, but with a large increase in effective area, longer baselines, and the possibility to form multiple independent station beams.

During standard astronomical observations, the station beams will be sent to a central correlator facility for image processing. For both pulsar timing and lunar observations, these station beams will be added in-phase to form `array beams', with up to $16$ dual polarisation array beams being produced for the phase $1$ instrument. Since the angular size of each array beam will be determined by the baseline over which it is formed, only the core stations will be used for lunar detection mode, so that the $16$ phase $1$ beams will cover approximately $50$\% of the lunar limb. These beams will be analysed in real-time by purpose-built hardware, which will perform the necessary processing to search for broad-bandwidth pulses \citep{lunar_technical}.
During phase $2$, a similar scheme is envisaged, but with sufficiently many array beams to cover the entire Moon.

\section{Simulating the SKA}
\label{sec:simulations}

The program used to simulate the sensitivity of the SKA was developed by \citet{JamesProtheroe09a}\footnote{Here, the simplified version of the simulation is used, which does not include secondary cascades produced by $\mu$ or $\tau$ from $\nu_{\mu}$ and $\nu_{\tau}$ charged-current interactions.}. It is a Monte Carlo routine that can simulate both neutrino and cosmic ray interactions in --- and in the former case, propagation through --- the Moon, and produces frequency-dependent radio emission from the resulting cascades according to parameterisations based on full simulations \citep{Alvarez-Muniz98,Alvarez-Muniz06}. The emission is propagated through a roughened lunar surface, and the signal spectrum as seen by a telescope on Earth is calculated, assuming no loss of coherence due to e.g.\ small-scale lunar surface roughness, or the ionosphere. Here, the former effect is ignored, since it is expected to be negligible for low-frequency cosmic ray observations, while the latter effect is assumed to be corrected for during signal processing. This program has been verified through comparison with analytic calculations of \citet{gayley2009}.

\begin{table*}
\begin{centering}
\begin{tabular}{l|ccccc}
	 &	$A_{\rm eff}/T_{\rm sys}$  & $f_{\rm min}$ & $f_{\rm max}$ & Beam coverage & $\sigma_{\rm thresh}$ \\
	 & 	m$^2$ K$^{-1}$ & MHz & MHz && \\
\hline
Phase 2 & $4,000$ & $100$ & $350$ & $100$\%& 10 \\
Phase 1 & $250$ & $100$ & $350$ & $\sim50\%$ & 7
\end{tabular}
\caption{Parameters of SKA-low used for sensitivity calculations: sensitivity $A_{\rm eff}/T_{\rm sys}$, minimum $f_{\rm min}$ and maximum $f_{\rm max}$ observation frequencies, characteristic lunar beam coverage, and detection threshold $\sigma_{\rm thresh}$.} \label{tab:ska}
\end{centering}
\end{table*}

The expected parameters of SKA-low phases $1$ and $2$ are given in Table \ref{tab:ska}. The limiting sensitivity of such an experiment is given when full-bandwidth data from all antennas are coherently added in-phase and used to search for Askaryan pulses over the entire lunar surface, assuming a perfect correction for ionospheric dispersion. Such an idealised analysis can be performed offline using buffered data. Hence, phase $2$ sensitivity is calculated assuming a detection threshold of $10 \sigma$ relative to the noise level of the entire array at full sensitivity, corresponding to a false trigger rate of much less than once per year.
However, the final sensitivity can also be determined by the real-time trigger, e.g.\ in the case of insufficient array beamforming capacity. This will be the case for phase $1$,  with sensitivity determined by the expected real-time trigger threshold of $7 \sigma$ (trigger rate of $\sim0.2$ Hz) using only the sensitivity of the core. The $16$ dual-linear polarisation array beams generated in real time are modelled assuming a Gaussian antenna density with fall-off radius of $240$~m, and $50\%$ of the total sensitivity of $500$~m$^2$~K$^{-1}$.
In both cases, the antenna noise calculation assumes a constant sensitivity over the full bandwidth, which approximates the effects of an increase in both effective area and sky noise at low frequencies, and an increased beam filling factor due to the Moon (which is colder than the sky at $100$~MHz, and hotter at $350$~MHz) at high frequencies. The frequency range $50$--$100$~MHz will not be used, due to the high sky noise and ionospheric dispersion

\section{Cosmic Rays}
\label{sec:cosmic_rays}

\begin{figure*}

\includegraphics[width=0.48\textwidth]{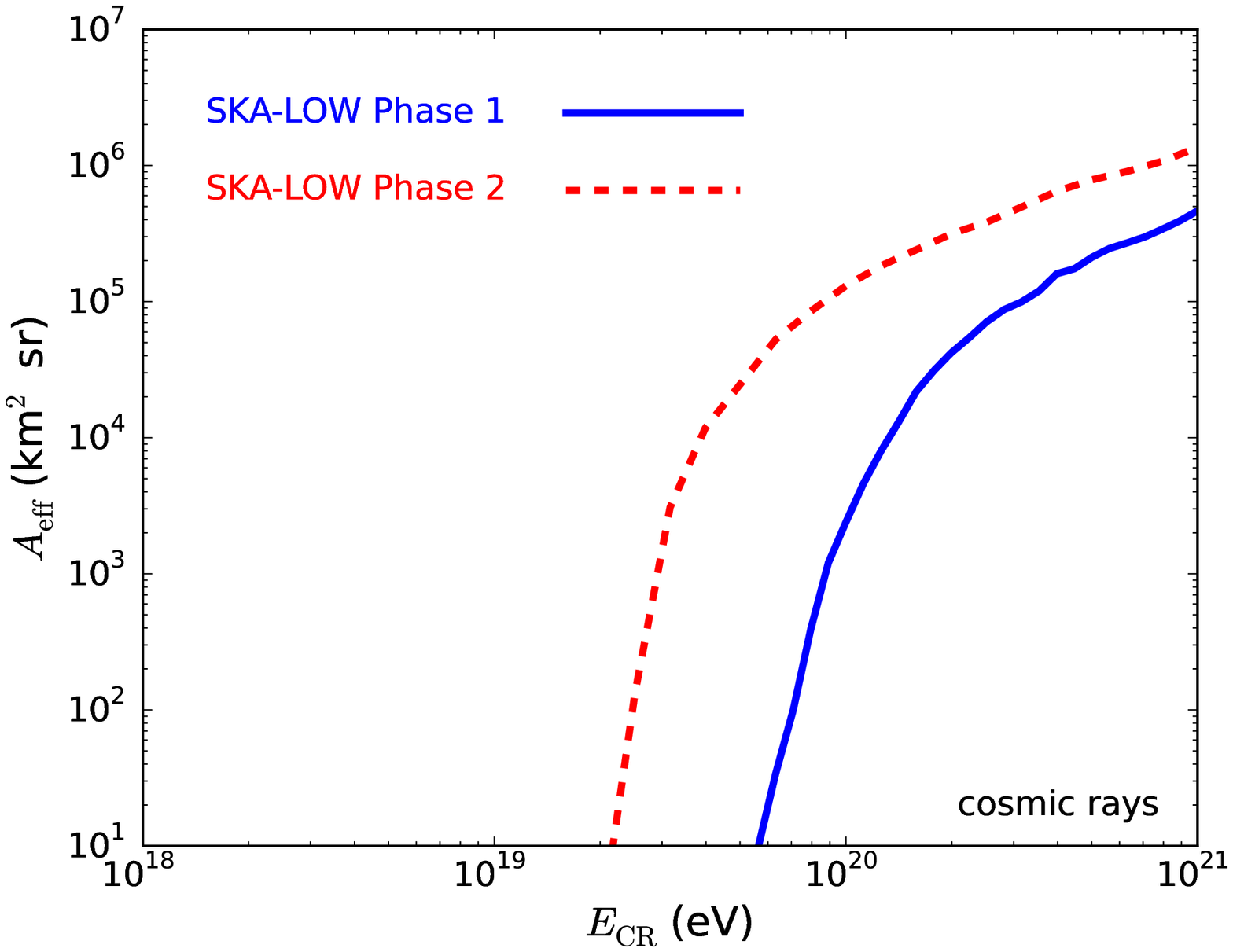} \includegraphics[width=0.48\textwidth]{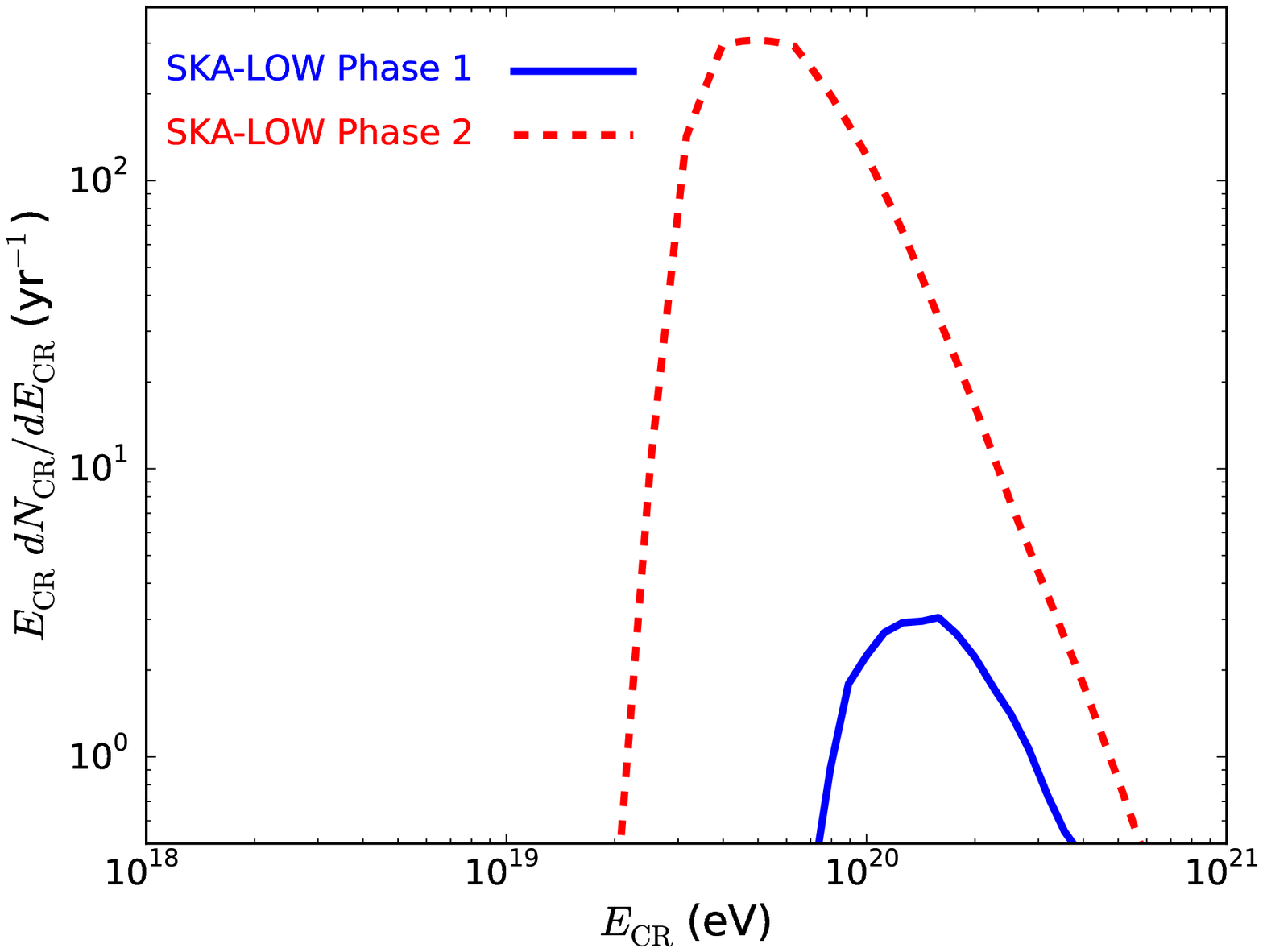}
\caption{Left: effective apertures of SKA-low phases $1$ and $2$ to cosmic rays as a function of their energy. Right: expected cosmic ray detection rates, after convolving with the measured flux \citep{Abraham2010}.} \label{fig:cr_apps}
\end{figure*}

The effective aperture of SKA-low to cosmic rays is given in Fig.\ \ref{fig:cr_apps} (left). The phase $1$ instrument achieves an instantaneous aperture of $10^{4}$~km$^2$~sr, comparable to that of the Pierre Auger Observatory with a typical $60^{\circ}$ zenith-angle cut \citep{Abraham2010}, at approximately $10^{20}$~eV, while phase $2$ does so at $\sim 3 \cdot 10^{19}$~eV. The relatively small reduction in threshold between phases $1$ and $2$ is due to the scaling of Askaryan emission: radiated power increases with the square of primary energy, so that a four-fold increase in sensitivity is required to halve the detection threshold. The rapid increase in $A_{\rm eff}$ above the detection threshold however means that at the very highest observed cosmic ray energies near $10^{20}$~eV, SKA-low phase $2$ will have an effective aperture of over $10^{5}$~km$^2$~sr.

The expected event rates per full year of observation time are shown in Fig.\ \ref{fig:cr_apps} (right), which have been calculated by convolving $A_{\rm eff}$ with the measured cosmic ray spectrum \citep{Abraham2010}, assuming a full year's worth of observations (not accounting for lunar visibility). Phase $1$ would be expected to see only a handful of events ($\sim 3$) per year, with the expected number being subject to uncertainties in the spectrum above $10^{20}$~eV. However, SKA phase $2$ will be sensitive to the cosmic ray flux above $2 \cdot 10^{19}$~eV, with an event rate of $\sim 355$ yr$^{-1}$, and more importantly, $\sim 171$ yr$^{-1}$ above $56$~EeV, where anisotropies in UHECR arrival directions have been observed \citep{PAO07C1,Auger_cena_2015}.

\subsection{Energy and directional resolution}

\begin{figure*}
\includegraphics[width=0.48\textwidth]{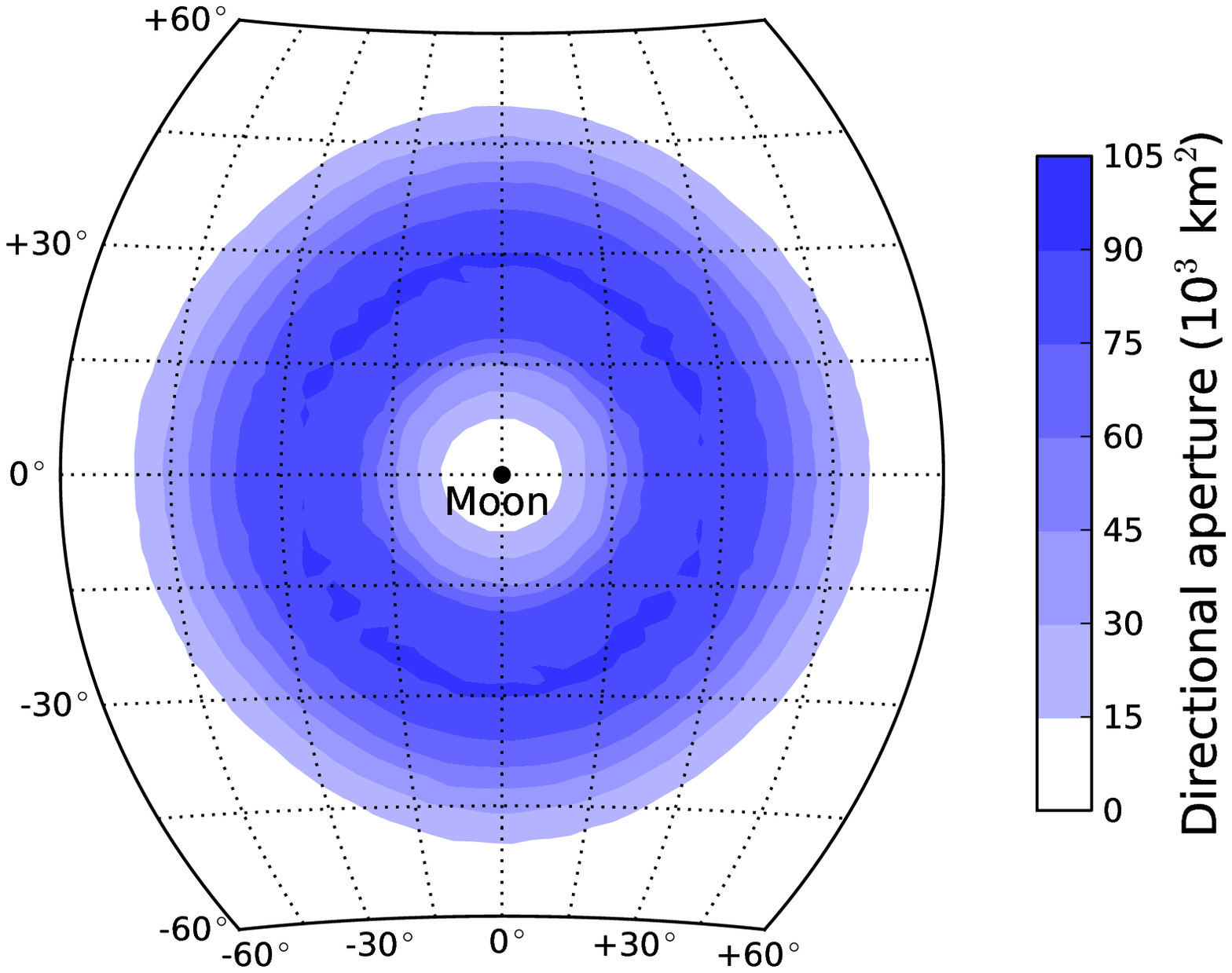} \includegraphics[width=0.48\textwidth]{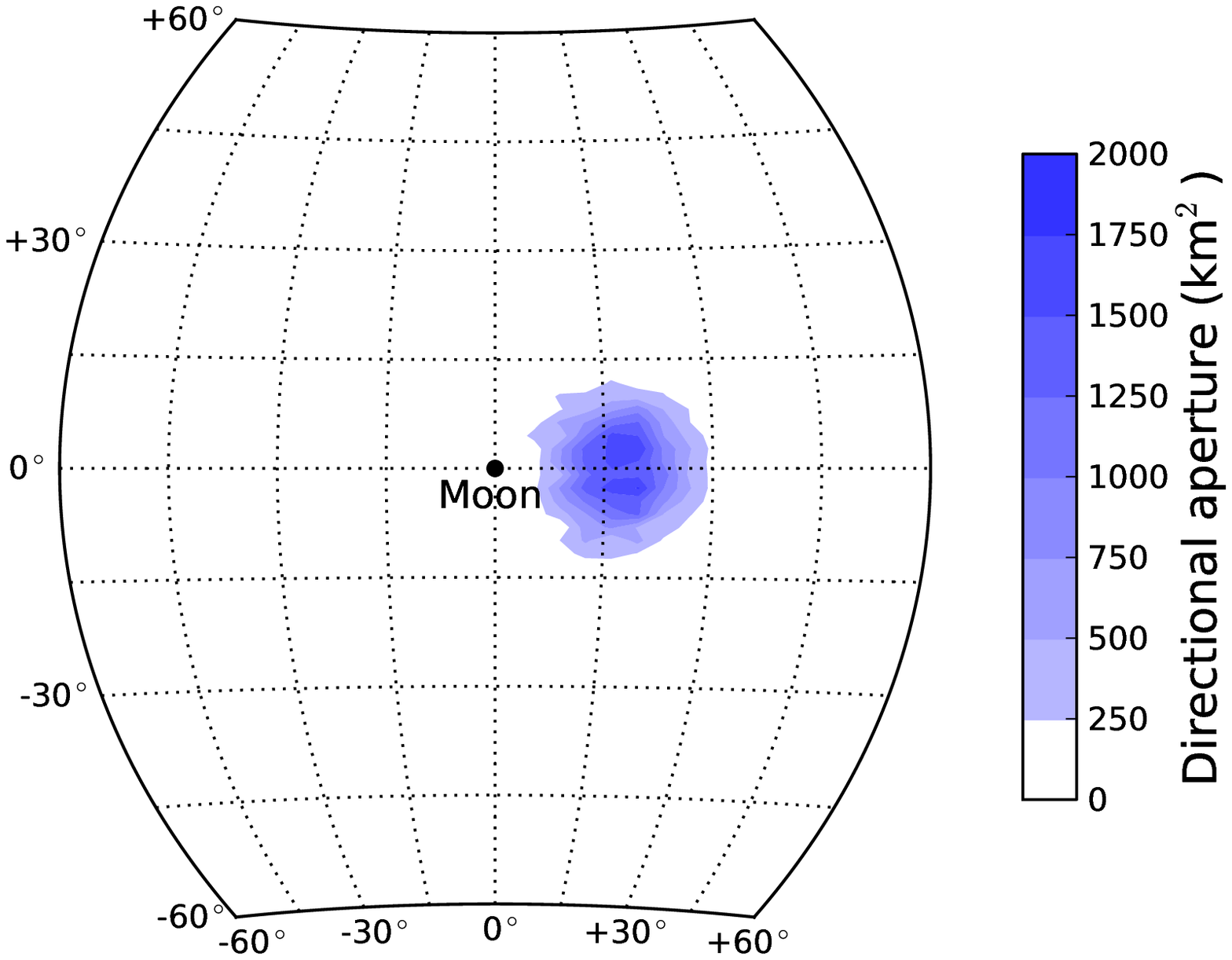}
\caption{Left: effective area of SKA-low phase $2$ to $10^{20}$~eV cosmic rays as a function of their arrival direction relative to the Moon. Right: the resulting effective area after placing cuts on the signal strength, origin, and polarisation (see text). The corresponding resolution for such cuts is $\sim5^{\circ}$.} \label{fig:ang_res}
\end{figure*}

Explicit reconstruction methods to determine the primary particle energy and arrival direction have not yet been developed for the lunar technique. Such methods would involve fitting the observed signal spectrum $\epsilon(f)$ over a broad bandwidth for a fully coherent component at low frequencies ($\epsilon(f) \propto f \, E_{\rm CR}$) to give the primary energy, measuring the (linear) polarisation to give the projection of the arrival direction in the plane of the sky, and fitting $\epsilon(f)$ for a spectral downturn at high frequencies to detect any decoherence due to observations away from the Cherenkov angle, in order to determine the angle of emission $\theta$ out of the plane of the sky.

A simple estimate of the minimum accuracy of this technique can be made by placing trivial cuts on the observable signal properties, and looking at the energy- and directional-dependence of the resulting reduced $A_{\rm eff}$ for these cuts.
The expected detection threshold of $10 \sigma$ gives an approximate error of $10$\% on the measured signal strength, and $0.1$~rad on the polarisation angle, which are modelled by only accepting signals with a magnitude in the $10$--$12\sigma$ range, and with a polarisation aligned within $\pm 5.7^{\circ}$ radially from the lunar limb. The expected angular resolution is taken as being $0.5'$, corresponding to only the inner part of the array, and only events originating from within $0.5'$ of a point on the lunar limb are accepted.

The result of this method is given in Fig.\ \ref{fig:ang_res}. The left-hand figure shows the effective area of SKA-low phase $1$ as a function of cosmic ray arrival direction relative to the Moon, for all detected events. The annulus can be interpreted as the instantaneous field of view of the SKA--Moon system to UHECR. On the right-hand side is plotted $A_{\rm eff}$ after cuts. The resulting acceptance has a characteristic resolution (given by the $1\sigma$ width) of $5^{\circ}$, centred on a region approximately $20^{\circ}$ from the centre of the Moon. Similar methods have been applied for the energy resolution, but while there are indications that SKA-low will have an energy resolution of better than $50$\%, this is mostly a function of the sharp detection threshold, and steeply falling cosmic ray spectrum \citep{2014arXiv1408.6069B}. It is expected that a specific reconstruction will be able to improve this accuracy significantly.

\section{Sensitivity to neutrinos}
\label{sec:neutrinos}

\begin{figure*}
\includegraphics[width=0.48\textwidth]{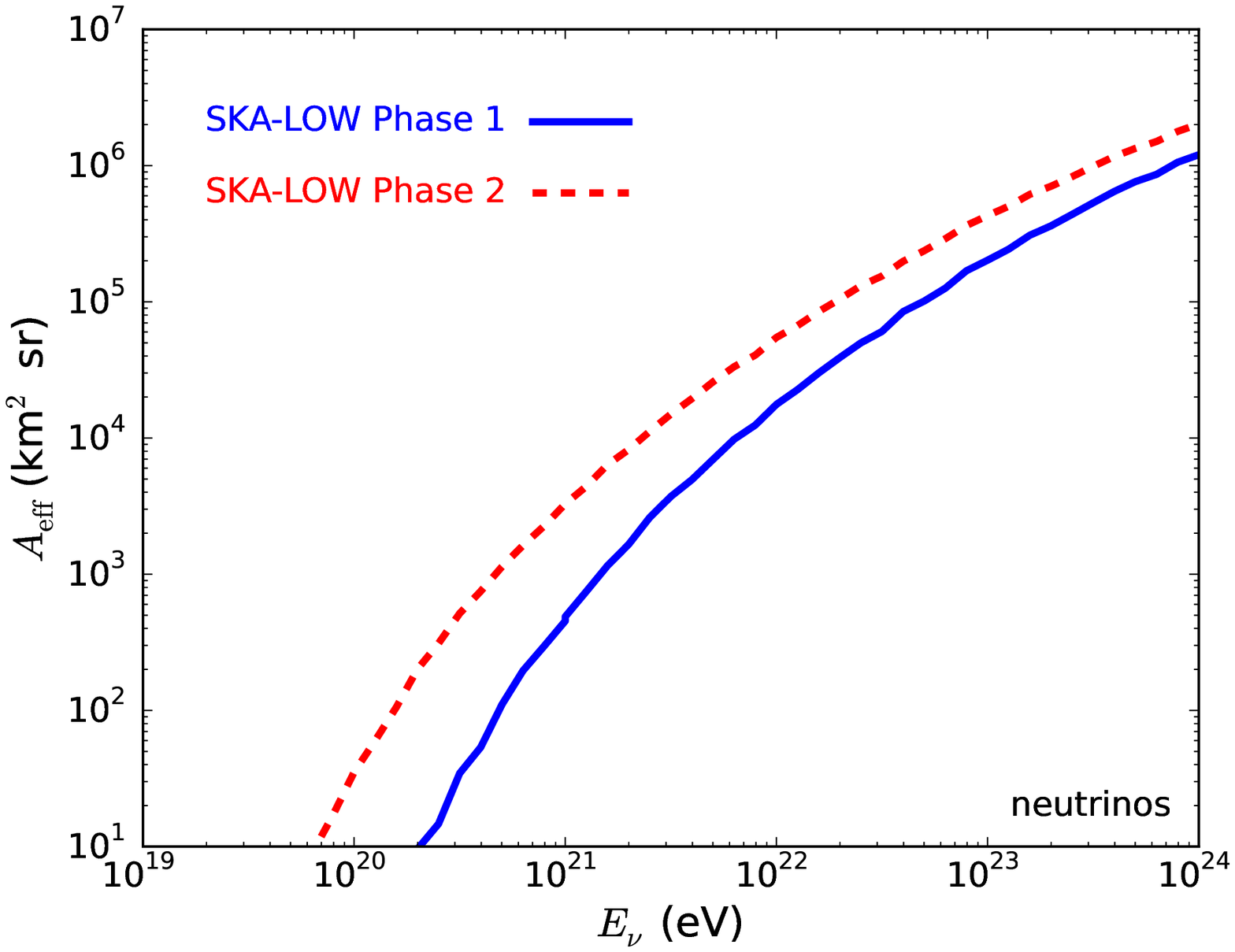} \includegraphics[width=0.48\textwidth]{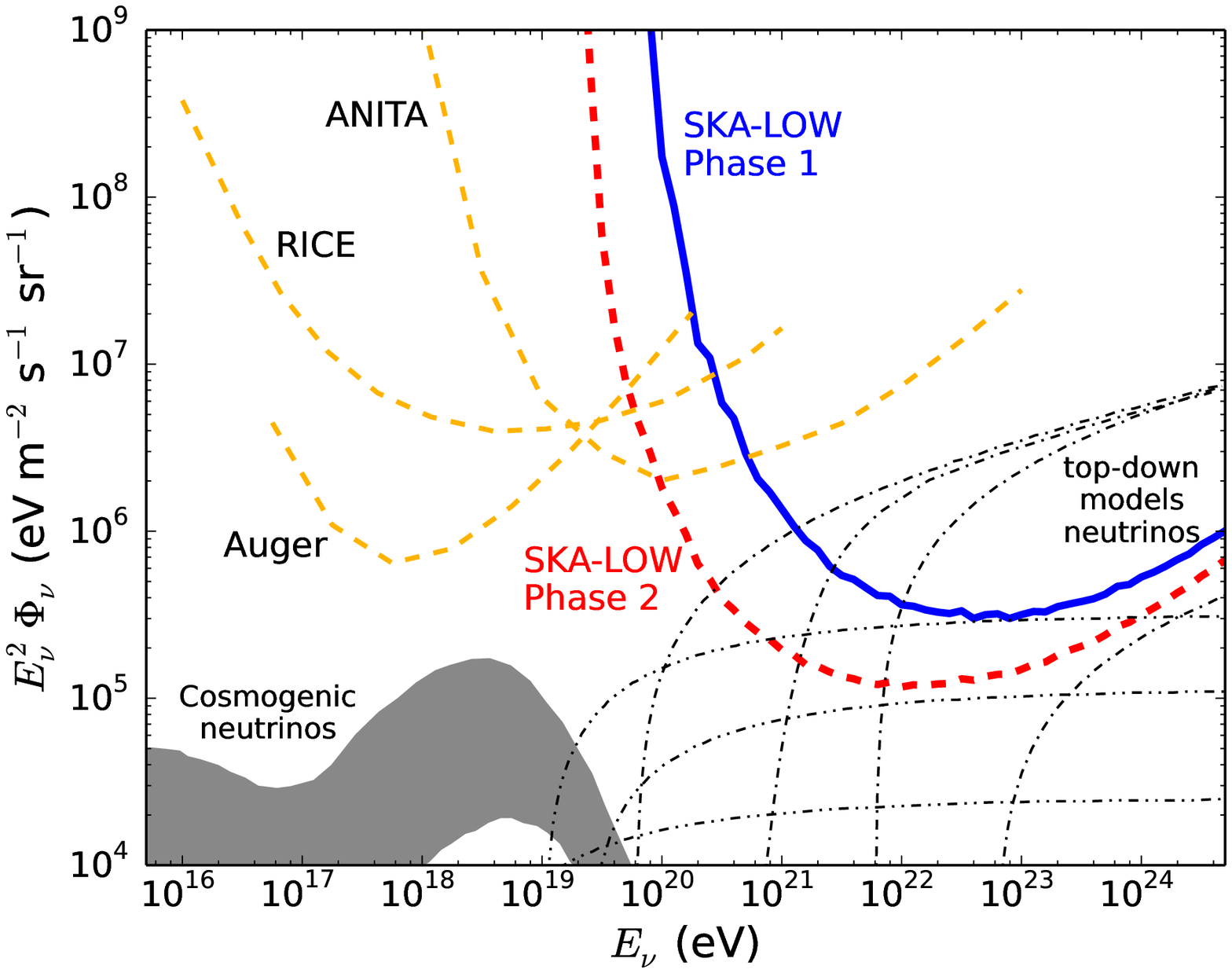}
\caption{Left: effective apertures of SKA-low phases $1$ and $2$ to UHE neutrinos as a function of their energy. Right: Projected $90\%$-confidence limits on the UHE neutrino flux from 1,000 hours of observations. Predictions are shown for neutrino fluxes from the so-called ``top-down models" involving the production of UHE neutrinos in the Early Universe from kinks (\citet{Lunardini12}, dash-dotted) and cusps (\citet{Berezinsky11}, dot-dash-dotted) in cosmic strings, and also for the neutrino flux 
produced in interactions of UHE cosmic-rays with the Cosmic Microwave Background radiation - ``cosmogenic neutrinos" (\citet{Allard06}, shaded). Limits set by other experiments - the Pierre Auger Observatory \citep{Auger_nu_2015}, RICE \citep{RICE} and ANITA \citep{ANITA,ANITAerratum} - are also shown. } \label{fig:nu_apps}
\end{figure*}

The instantaneous effective apertures $A_{\rm eff}$ to UHE neutrinos of SKA-low phases $1$ and $2$ are shown in Fig.\ \ref{fig:nu_apps}. The apertures are much smaller than that to cosmic rays, both because only $20\%$ of the primary neutrino energy is characteristically given to hadronic cascades, and because the majority of neutrinos interact too deeply in the Moon to be detected (field absorption length at $100$~MHz is of the order of $200$~m \citep{reg_absorption}. Therefore, an appreciable sensitivity is not reached until well above $10^{20}$~eV, so that the SKA will not be sensitive to neutrinos from the GZK process. However, the SKA will be able to probe ``top-down'' models of cosmic ray production, as shown in Fig.\ \ref{fig:nu_apps} (right) for a nominal $1000$~hr observation. While the projected sensitivity of ARA \citep{2012APh....35..457A} and ARIANNA \citep{2014arXiv1410.7352A} are not shown, their target sensitivity is to the GZK flux around $10^{18}$~eV, and is not expected to be competitive with SKA-low phase 2 at energies much above $10^{20}$~eV.

\section{Discussion}
\label{sec:discussion}

SKA-low phase 2, using the lunar Askaryan technique, is expected to have an instantaneous effective aperture of approximately $10^{5}$~km$^2$~sr to cosmic rays at $10^{20}$~eV. This gives it the potential to detect an unprecedented number of UHECR. Even folding in the lunar visibility of $29\%$ (assumed elevation limit of $30^{\circ}$) from the site of the SKA-low core in the Murchison Radio Astronomy Observatory in Western Australia, the expected annual detection rate of $\sim50$ cosmic rays with energy above $56$~EeV gives a method to increase detection statistics beyond that of the current Pierre Auger Observatory \citep{Auger_cena_2015} and Telescope Array samples \citep{abbasi2014b}. The only proposed experiments which could rival this detection rate are JEM-EUSO, to be placed on-board the international space station \citep{takahashi2009}, and the LORD mission \citep{2013JPhCS.409a2096R}.

The angular and energy resolutions of observations with SKA phase $2$ will not be competitive with those of current ground-based experiments, although they are expected to improve beyond the simplistic estimates of $5^{\circ}$ and $50$\% respectively presented here. The angular resolution however is already comparable to, or better than, the angular deflection expected from galactic magnetic fields \citep{dolag05,armengaud05}, and would certainly be able to test the observed excess in the $\sim15^{\circ}$ surrounding Centaurus A \citep{Auger_cena_2015}. In general, the SKA will be sensitive to objects lying within $\sim 45^{\circ}$ of the lunar orbit, i.e.\ more than half the sky, in a band centred on the Moon's orbit. For SKA-low, this band also includes the Galactic Centre and M87.

In order to compete with dedicated experiments such as JEM-EUSO or LORD (or, indeed, the $10$-year exposure of Pierre Auger), it will be vital to observe the Moon with SKA phase 2 whenever it is visible. This will require implementing a fully commensal observation mode, so that lunar observations will not compete with the many other science goals of the SKA.
Such a mode will be feasible for SKA phase 2, due to the planned ability to form multiple beams per station (see Sec.\ \ref{sec:ska}).  For SKA phase $1$ however, there will be more competition for the available beams, so the target is to perform engineering studies for phase $2$, and observations targeting models of the UHE neutrino flux. For each phase, specialised pulse detection hardware will be built, analogous to the `Bedlam Board' produced for the LUNASKA experiments at Parkes \citep{2013ExA....36..155B}, and the `ultra-high-energy particles' (UHEP) mode of LOFAR developed by the NuMoon collaboration \citep{singh2012} --- for further details, see the co-contribution by \citet{lunar_technical}.

\section{Conclusion}

The low-frequency component of the SKA will be a powerful instrument for performing ultra-high-energy particle astronomy. In order to enable these observations with a giant radio array, members of previous lunar experiments have joined forces to form the SKA High Energy Cosmic Particles focus group, together with an experiment to perform precision measurements of EAS. Using the lunar Askaryan technique, phase $1$ of SKA-low will be able to test models of top-down particle production, while phase $2$ will be able to detect an unprecedented number of cosmic rays at energies above $56$ EeV where indications of anisotropy are observed.

\vspace{0.5cm}

\noindent {\bf acknowledgement}

The authors would like to dedicate this contribution to their friend, colleague, and fellow `lunatic' Ray Protheroe, who sadly passed away on Wednesday July $1^{\rm st}$, 2015. As one of the first to promote the lunar technique, and as founding member of the LUNASKA project, he is in large part responsible for the existence of the HECP group.

\let\OLDthebibliography\thebibliography
\renewcommand\thebibliography[1]{
  \OLDthebibliography{#1}
  \setlength{\parskip}{0pt}
  \setlength{\itemsep}{0pt plus 0.3ex}
}


\begin{thebibliography}{99}

\begin{small}

\bibitem[Aab et al.(2015)a]{Auger_nu_2015}
Aab, A.\ {\it et al.} (Pierre Auger Collaboration), 2015,
Phys.Rev.D, {\bf 91}, 092008.

\bibitem[Aab et al.(2015)b]{Auger_cena_2015} Aab, A.\ {\it et al.} (Pierre Auger Collaboration), 2010, ApJ, {\bf 804}, 15.

\bibitem[Abbasi et al.(2014)]{abbasi2014b} Abbasi, R.U.\ et al.\ (Telescope Array Coll.), 2014, ApJ, {\bf 790}, L21 

\bibitem[Abraham et al.\ (2010)]{Abraham2010} Abraham, J.\ et al., 2010, Phys.Lett.B, {\bf 685}, 239.

\bibitem[Abraham et al.(2007)]{PAO07C1}
   Abraham, J.\ et al.\ (Pierre Auger Coll.), 2007, Science, {\bf 318}, 938

\bibitem[Allard et al.(2006)]{Allard06} Allard, D., Ave, M., Busca, N.\ et al., 2006, JCAP, {\bf 9}, 005.

\bibitem[Allison et al.(2012)]{2012APh....35..457A} Allison, P.\ et al.\ (ARA Collaboration), 2012, Astropart.Phys., {\bf 35}, 457. 

\bibitem[Alvarez-Mu\~{n}iz et al.(2006)]{Alvarez-Muniz06} Alvarez-Mu\~{n}iz, J., Marqu\'es, E., V\'azquez, R.A., \& Zas, E., 2006, Phys.Rev.D, {\bf 74}, 023007

\bibitem[Alvarez-Mu\~{n}iz \& Zas(1998)]{Alvarez-Muniz98} Alvarez-Mu\~{n}iz, J., Zas, E., 1998,  Phys.Lett.B, {\bf 434}, 396. 


\bibitem[Armengaud, Sigl \& Miniati(2005)]{armengaud05} Armengaud, E., Sigl, G., \& Miniati, F., 2005, Phys.Rev.D, {\bf 72}, 043009.

\bibitem[Askaryan(1962)]{askaryan1962} Askaryan, G.A., 1962, Sov.\ Phys.\ JETP, {\bf 14}, 441

\bibitem[Barwick et al.(2014)]{2014arXiv1410.7352A} Barwick, S.W.\ et al.\ (ARIANNA Collaboration), 2014, arXiv:1410.7352. 

\bibitem[Beresnyak et al.(2005)]{beresnyak2005} Beresnyak, A.R., Dagkesamanskii, R.D., \& Zheleznykh, I.M.\ et al., 2005, Astron.Rep., {\bf 49}, 127

\bibitem[Berezinsky et al.(2011)]{Berezinsky11} Berezinsky, V., Sabancilar, E.\ \& Vilenkin, A., 2011, Phys.Rev.D, {\bf 84}, 085006

\bibitem[Bray et al.(2015)a]{2015APh....65...22B} Bray, J.D.\ et al., 2015, APh, {\bf 65}, 22.

\bibitem[Bray et al.(2015)b]{2014arXiv1408.6069B}
Bray, J.D.\ et al., 2015, aska.conf, {\bf 144} [arXiv:1408.6069B].

\bibitem[Bray et al.(2015)c]{lunar_technical} Bray, J.D.\ et al., 2015, \emph{these proceedings} (I.D.\ 597).

\bibitem[Bray, Ekers \& Roberts(2013)]{2013ExA....36..155B} Bray, J., Ekers, R.D., \& Roberts, P., Exp.Astronomy, {\bf 36}, 155.

\bibitem[Buitink et al.(2010)]{buitink2010} Buitink, S.W.\ et al., 2010, A\&A, {\bf 521}, A47

\bibitem[Dagkesamanskii \& Zheleznykh(1989)]{dagkesamanskii1989} Dagkesamanskii, R.D.\ \& Zheleznykh, I.M., 1989, Sov.Phys.JETP Lett., {\bf 50}, 259.

\bibitem[Dewdney et al.(2013)]{SKA_baseline_design} Dewdney, P.E.\ et al., 2013, ``SKA1 System Baseline Design'', \mbox{SKA-TEL-SKO-DD-001}

\bibitem[Dolag et al.(2005)]{dolag05} Dolag, K.\ et al., 2005, JCAP, {\bf 0501}, 009.

\bibitem[Gayley et al.(2009)]{gayley2009} Gayley, K.G., Mutel, R.L.\ \& Jaeger, T.R., 2009, ApJ, {\bf 706}, 1556

\bibitem[Gorham et al.(2004)]{gorham2004a} Gorham, P.W.\ et al., 2004, Phys.Rev.Lett., {\bf 93}, 041101

\bibitem[Gorham et al.(2010)]{ANITA} Gorham, P.W.\ et al.\ (ANITA Coll.), 2010, Phys.Rev.D, {\bf 82}, 022004

\bibitem[Gorham et al.(2012)]{ANITAerratum} Gorham, P.W.\ et al.\ (ANITA Coll.), 2012, Phys.Rev.D, {\bf 85}, 049901

\bibitem[Huege et al.(2015)b]{ska_eas} Huege, T.\ et al., 2015, \emph{these proceedings} (I.D.\ 309).

 \bibitem[Jaeger et al.(2010)]{jaeger2010} Jaeger, T.R., Mutel, R.L.\ \& Gayley, K.G., 2010, Astropart.Phys., {\bf 43}, 293

 \bibitem[James \& Protheroe(2009)]{JamesProtheroe09a} James, C.W.\ \& Protheroe, R.J., 2009, Astropart.Phys., {\bf 30}, 318

 \bibitem[James et al.(2010)]{james2010} James, C.W.\ et al., 2010, Phys.Rev.D, {\bf 81}, 042003

\bibitem[Kravchenko et al.(2012)]{RICE} Kravchenko, I.\ et al., 2012, Phys.Rev.D, {\bf 85}, 062004.

 \bibitem[Lunardini \& Sabancilar(2012)]{Lunardini12} Lunardini, C. \& Sabancilar, E., 2012, Phys.Rev.D, {\bf 86}, 085008

\bibitem[Olhoeft \& Strangway(1975)]{reg_absorption} Olhoeft, G.\ \& Strangway, D., 1975, Earth Planet.Sci.Lett., {\bf 24}, 394.

\bibitem[Ryabov, Gusev \& Chechin(2013)]{2013JPhCS.409a2096R} Ryabov, V.A., Gusev, G.A.\ \& Chechin, V.A., 2013, J.Phys.Conf.Series, {\bf 409}, 012096. 

\bibitem[Scholten et al.(2006)]{NuMoonSim} Scholten, O.\ et al., 2006, Astropart.Phys, {\bf 26}, 219.

\bibitem[Singh et al.(2012)]{singh2012} Singh, K.\ et al.\ (LOFAR Coll.), 2012, NIMA, {\bf 664}, 171.

\bibitem[Spencer et al.(2010)]{spencer2010} Spencer, R.E., Macfarlane, A., Mills, O.\ \& Piccirillo, L., 2010, Proc.\ 10$^{\rm th}$ EVN Symposium, 097.

\bibitem[Takahashi et al.(2009)]{takahashi2009} Takahashi, Y.\ et al.\ (JEM-EUSO Coll.), 2009, New J.Phys., {\bf 11}, 065009.

\end{small}
\end{thebibliography}
\end{document}